\documentclass[showpacs,aps,prb,twocolumn,floatfix]{revtex4-1}
\usepackage{amsmath}
\usepackage{amsfonts}
\usepackage{amssymb}
\usepackage{amsthm}
\usepackage{graphicx}
\usepackage{color}
\usepackage[verbose,hypertexnames=false,bookmarksopenlevel=1,filecolor=blue,
linkcolor=blue,citecolor=blue,pdfstartview=FitH,bookmarksopen,bookmarksnumbered,
colorlinks,plainpages=false,linktocpage]{hyperref}

\definecolor{orange}{rgb}{1,0.5,0}

\DeclareMathOperator\artanh{artanh}
\DeclareMathOperator\diag{diag}
\DeclareMathOperator\sign{sign}
\begin{document}
\title{Trigonal Warping in Bilayer Graphene: 
Energy versus Entanglement Spectrum}
\author{Sonja Predin, Paul Wenk, and John Schliemann}
\affiliation{ Institute for Theoretical Physics, University of
Regensburg, D-93040 Regensburg, Germany}
\date{December 2015}

\begin{abstract}
  We present a mainly analytical study of the entanglement spectrum of
  Bernal-stacked graphene bilayers in the presence of trigonal warping
  in the energy spectrum. Upon tracing out one layer, the entanglement
  spectrum shows qualitative geometric differences to the energy
  spectrum of a graphene monolayer. However, topological quantities
  such as Berry phase type contributions to Chern numbers agree. The
  latter analysis involves not only the eigenvalues of the
  entanglement Hamiltonian but also its eigenvectors.
 
  We also discuss the entanglement spectra resulting from tracing out
  other sublattices. As a technical basis of our analysis we provide
  closed analytical expressions for the full eigensystem of bilayer
  graphene in the entire Brillouin zone with a trigonally warped
  spectrum.
\end{abstract}
\maketitle
\section{Introduction}
Although first considered as a source of quantum corrections to the
entropy of black holes\cite{PhysRevD.34.373}, entanglement entropy, in
particular the von Neumann entropy, evolved into a tool in the field
of many-body systems. This brought along connections between seemingly
unrelated research areas. In condensed matter, the entanglement
entropy serves, e.g., as a geometrical interpretation for the boundary
between local quantum many-body systems. This connection has its
origin in the \textit{area laws}\cite{RevModPhys.82.277}.

However, Li and Haldane have shown that the related entanglement
spectrum contains more information than the single number expressed by
the entanglement entropy\cite{Li08}. 
This spectrum is determined by the Schmidt decomposition of the ground
state of a bipartite system, and the reduced density matrix obtained
by tracing out one of the subsystems can always be formulated as
\begin{equation}
\rho_{\rm red}=\frac{e^{-{\cal H}_{\rm ent}}}{Z}
\label{rhoentham}
\end{equation}
with an entanglement Hamiltonian ${\cal H}_{\rm ent}$ encoding the
entanglement spectrum, and a partition function $Z={\rm tr}(e^{-{\cal
    H}_{\rm ent}})$. Following the Li-Haldane conjecture\cite{Li08},
in a gaped phase, the entanglement spectrum can be directly related to
the spectrum of edge excitations as shown for the Fractional Quantum
Hall System\cite{Lauchli10, Thomale10, Chandran11}.  This relation to
the edge excitations can be also seen analytically in the case of
non-interacting particles. It can be shown by mapping the free
fermionic system ${\cal H}$ onto a \textit{flat-band} Hamiltonian
${\cal H}_{\text{flat}}$\cite{Kitaev06}. Now, the eigenenergies $e_i$
of the latter are related to the eigenenergies of the corresponding
entanglement energies $\varepsilon_i$ as $e_{i}\sim
\tanh\left(\varepsilon_{i}/2\right)/2+\text{const}.$\cite{Turner10}
Thereby, the eigenstates of both ${\cal H}_{\text{flat}}$ and ${\cal
  H}$ are the same. Thus, if ${\cal H}$ contains topologically
protected surface states the same holds for the entanglement
Hamiltonian.

This is why the entanglement spectrum, beyond the related entropy, is
considered a \textit{tower of states} and used as a fingerprint for
topological order. However, this is not true in general as shown
recently by A. Chandran \textit{et al.},
Ref.~\onlinecite{PhysRevLett.113.060501}.

As a result of a multitude of studies, there is a plethora of
revisited effects in the context of entanglement spectrum like the
Kondo effect, many-body localization or disordered quantum spin
systems; for recent reviews see
Refs.~\onlinecite{Regnault15,Laflorencie15}.

A particular situation
arises if the edge comprises the entire remaining subsystem as it is
the case for spin ladders \cite{Poilblanc10, Cirac11, Peschel11,
  Lauchli11, Schliemann12, Tanaka12, Lundgren12, Lundgren13, Chen13,
  Lundgren14} and various bilayer systems
\cite{Schliemann11,Schliemann13,Schliemann14}. A typical observation
in such scenarios is, in the regime of strongly coupled subsystems, a
proportionality between the energy Hamiltonian of the remaining
subsystem and the appropriately defined entanglement Hamiltonian. We
note that the entanglement Hamiltonian entering the reduced density
matrix (\ref{rhoentham}) is only determined up to multiples of the
unit operator which has consequences regarding thermodynamic relations
between the entanglement entropy and the subsystem energy
\cite{Schliemann11,Schliemann13,Schliemann14}.

On the other hand, such a close relation between energy and
entanglement Hamiltonian is not truly general as shown in
Ref.~\onlinecite{Lundgren12} where a spin ladder of clearly
nonidentical legs was studied. In the present work we provide another
counter example given by graphene bilayers in the presence of trigonal
warping \cite{McCann13,Rozhkov15}. As we shall see in the following,
the geometric properties of the the entanglement spectrum of an
undoped graphene bilayer and the energy spectrum of a monolayer
clearly differ qualitatively.  However, certain topological quantities
such as Berry phase type contributions to Chern numbers agree. The
latter analysis involves not only the eigenvalues of the entanglement
Hamiltonian (i.e., the entanglement spectrum) but also its
eigenvectors.

This paper is organized as follows. In section \ref{energy} we discuss
the full eigensystem of the tight-binding model of bilayer graphene in
the presence of trigonal warping; a full account of the technical
details is given in appendices \ref{diag} and \ref{continuity}. To
enable analytical progress we neglect here terms breaking
particle-hole symmetry.  On the other hand, our calculation considers
the entire first Brillouin zone and avoids the Dirac cone
approximation usually employed in studies of trigonal warping in
graphene bilayers \cite{McCann06, Nilsson06, Koshino06, Kechedzhi07,
  Manes07, Cserti07, Mikitik08, Mariani12, Cosma15}.  We compare our
results for the full four-band model with an effective Hamiltonian
acting on the two central bands \cite{McCann06,Mariani12,Cosma15}.
The entanglement spectrum obtained from the ground state of undoped
graphene bilayers is analyzed in section \ref{entanglement}. We
discuss the case of one layer being traced out as well as the
situation where the trace is performed over two other out of four
sublattices.  We close with a summary and an outlook in section
\ref{concl}.

\section{Energy Spectrum of Graphene Bilayers: Trigonal Warping and
  Topological Invariants}
\label{energy}
\allowdisplaybreaks
The standard tight-binding Hamiltonian for graphene bilayers in Bernal
stacking can be formulated as \cite{McCann13,Rozhkov15}
\begin{align}
  H = {}& -t\sum_{\vec k} \left(\gamma(\vec k)a_{1\vec k}^\dagger b_{1\vec
      k} + \gamma(\vec k)a_{2\vec k}^\dagger b_{2\vec k}
    +{\rm h.c.}\right)\nonumber\\
  {}& +t_{\perp}\sum_{\vec k}\left(b_{1\vec k}^\dagger a_{2\vec k}
    +a_{2\vec k}^\dagger b_{1\vec k}\right)\nonumber\\
  {}& -t_3\sum_{\vec k}\left(\gamma(\vec k)b_{2\vec k}^\dagger a_{1\vec k}
    +\gamma^{\ast}(\vec k)a_{1\vec k}^\dagger b_{2\vec k}\right)\nonumber\\
  {}& +t_4\sum_{\vec k}\left(\gamma(\vec k) \left(a_{1\vec
        k}^\dagger a_{2\vec k}+b_{1\vec k}^\dagger b_{2\vec k}\right)+{\rm
      h.c.}\right)\,,
\label{hambilayer}
\end{align}
where $a_{i\vec k}^\dagger $ ($a_{i\vec k}$) and $b_{i\vec k}^\dagger $ ($b_{i\vec
  k}$) create (annihilate) electrons with wave vector $\vec k$ in
layers $i=1,2$ on sublattice A and B, respectively. Moreover,
$\gamma(\vec k)=\sum_{l=1}^3\exp(i\vec k \cdot \vec\delta_l)$ where the
$\vec\delta_l$ are the vectors connecting a given carbon atom with its
nearest neighbors on the other sublattice in a graphene monolayer.  In
what follows we will use coordinates with
\begin{equation}
  \vec\delta_{1,2}=\frac{a}{2}\left(-1,\pm\sqrt{3}\right)\quad,\quad
  \vec\delta_3=a(1,0)
\label{defdelta}
\end{equation}
where $a=1.42$\AA\, is the distance between neighboring carbon atoms,
such that the two inequivalent corners of the first Brillouin zone can
be given as
\begin{equation}
{\vec K}_{\pm}=\frac{2\pi}{3\sqrt{3}a}\left(\sqrt{3},\pm 1\right).
\label{defK}
\end{equation}
The parameter $t$ describes hopping within each layer between the
sublattices while $t_{\perp}$ parameterizes the vertical hopping
between the two sublattices in different layers lying on top of each
other. The additional hopping processes described by the skew
parameters $t_3$, $t_4$ lead to trigonal warping of the spectrum and
electron-hole asymmetry, respectively.  Experimentally established
values \cite{Kuzmenko09} for these quantities are $t=3.16{\rm eV}$,
$t_{\perp}=0.381{\rm eV}$, $t_3=0.38{\rm eV}$, and $t_4=0.14{\rm eV}$.
\begin{figure}[t]
\includegraphics[width=0.7\columnwidth]{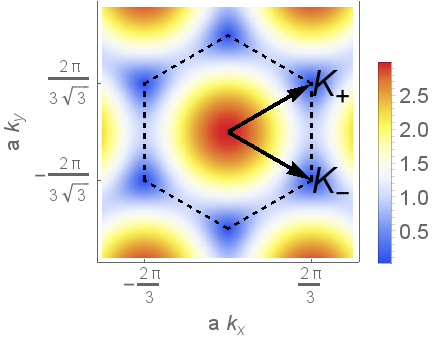}
\caption{(Color Online)
 Brillouin zone with a density plot of $|\gamma(\vec k)|$.}
\label{fig1}
\end{figure}
The geometry of the first Brillouin zone is visualized in Fig.~\ref{fig1}
along with a color plot of the modulus $|\gamma(\vec k)|$.

\begin{figure}[t]
  \includegraphics[width=\columnwidth]{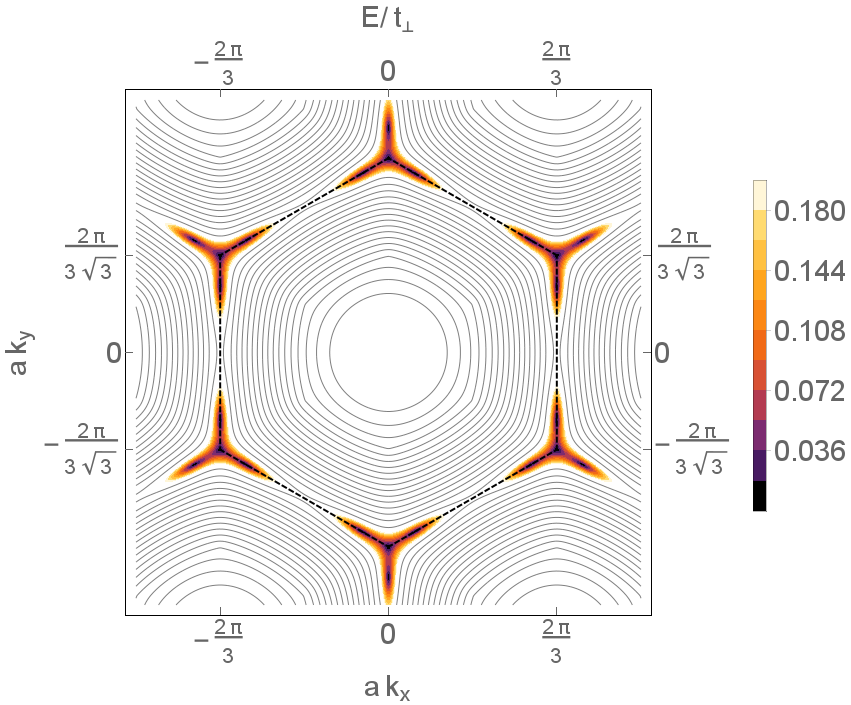}
  \caption{(Color Online) 
    Contour plot of the energy band $(+E_2(\vec k))$ plotted for
    $t_{\perp}=t$, $t_3=0.5t$. The contour of the colored region
    indicates $E=0.2/t_\perp$. The edge of the first Brillouin zone is marked
    by dashed lines.}
\label{fig2}
\end{figure}
The presence of all four couplings in the Hamiltonian
Eq.~(\ref{hambilayer}) makes its explicit diagonalization in terms of
analytical expressions a particularly cumbersome task. As the present
study chiefly relies on analytical calculations rather than resorts to
numerics, we will drop the contributions proportional to the smallest
parameter $t_4$ in order to achieve an analytically manageable
situation.

\begin{figure}[t]
\includegraphics[width=\columnwidth]{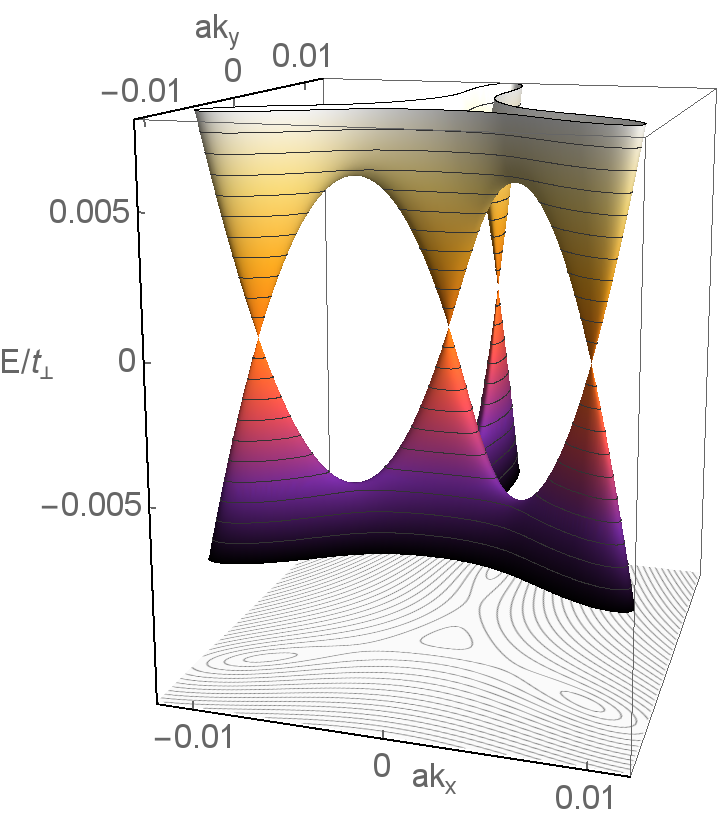}
\caption{(Color Online) 
  The central energy bands $(\pm E_2(\vec k))$ plotted around a
  given $K$-point for $t_{\perp}=0.1t$, $t_3=0.15t$. The dispersions
  show a central Dirac cone accompanied by three satellites.
  The components of the wave vector are measured relatively to the $K$-point.}
\label{fig3}
\end{figure}

Putting $t_4=0$ the full eigensystem of the Hamiltonian
(\ref{hambilayer}) can be obtained in a closed analytical fashion as
detailed in appendix \ref{diag}.  The four dispersion branches $(\pm
E_1(\vec k))$, $(\pm E_2(\vec k))$ form a symmetric spectrum with
\begin{small}
\begin{widetext}
\begin{equation}
  E_{1/2}=\sqrt{\frac{1}{2}\left(t_{\perp}^2+t_3^2|\gamma(\vec k)|^2
      +2t^2|\gamma(\vec k)|^2
      \pm\sqrt{4t^2|\gamma(\vec k)|^2\left(t_{\perp}^2+t_3^2|\gamma(\vec k)|^2
          -2t_{\perp}t_3|\gamma(\vec k)|\cos\left(3\phi_{\vec k}\right)\right)
        +\left(t_{\perp}^2-t_3^2|\gamma(\vec k)|^2\right)^2}\right)}
\label{ergspec}
\end{equation}
\end{widetext}
\end{small}
and $\gamma(\vec k)=|\gamma(\vec k)|e^{i\phi_{\vec k}}$. The two outer
branches $(\pm E_1(\vec k))$ are separated from the inner ones $(\pm
E_2(\vec k))$ by gaps determined essentially by the hopping parameter
$t_{\perp}$.  The result Eq.~(\ref{ergspec}) generalizes the energy
spectrum given in Ref.~\onlinecite{McCann06} within the Dirac cone
approximation to the full Brillouin zone. Moreover, in appendix
\ref{diag} we also give the complete data of the corresponding
eigenvectors. Fig.~\ref{fig3} concentrates on the vicinity of a given
$K$-point using realistic parameters.

The inner branches $(\pm E_2(\vec k))$ dominate the low-energy physics of the 
system near half filling and meet at zero energy for
\begin{equation}
  \gamma(\vec k)=0
\label{cond1}
\end{equation}
corresponding to the two inequivalent corners $K_{\pm}$ of the first
Brillouin zone, and for
\begin{equation}
  \cos\left(3\phi_{\vec k}\right)=-1\quad
  \wedge\quad|\gamma(\vec k)|=\frac{t_{\perp}t_3}{t^2}\,.
\label{cond2}
\end{equation}
The latter condition defines three additional satellite Dirac cones
around each $K$-point two of which lying on the edges (faces) of the
Brillouin zone connecting $K_{\pm}$. The third satellite Dirac cone
lies formally outside the Brillouin zone but is equivalent to a
satellite cone on the edge around an equivalent $K$-point. Indeed, the
quantity $\gamma(\vec k)$ has a constant phase $\phi_{\vec
  k}\in\{-\pi/3,\pi/3,\pi\}$ on each face: As an example, consider the
edge connecting the two inequivalent $K$-points given in
Eq.~(\ref{defK}) where one finds
\begin{equation}
  \gamma\left(\frac{2\pi}{3a},k_y\right)
  =e^{-i\pi/3}\left(2\cos\left(\frac{\sqrt{3}}{2}k_ya\right)-1\right)
\end{equation}
with the parenthesis being nonnegative for $k_y$ ranging between 
$(\pm2\pi/(3\sqrt{3}a))$. Thus, solving for $k_y$ the satellite Dirac cones
on that edge lie at
\begin{equation}
  \vec k=\left(\frac{2\pi}{3a},
    \pm\frac{2}{\sqrt{3}a}
    \arccos\left(\frac{1}{2}\left(1+\frac{t_{\perp}t_3}{t^2}\right)\right)\right)\,,
\end{equation}
and the other satellite cones are located at positions being
equivalent under reciprocal lattice translation and/or hexagonal
rotation.  Note that for $t_{\perp}t_3/t^2=1$ the satellite cones
merge in the $M$-points (centers of the faces) and they vanish for
even larger values of that ratio.  In Fig.~\ref{fig2} we give a sketch
of the situation in the entire Brillouin zone for moderate values of
$t_{\perp}$ and $t_3$.
For $t_3=0$ the two energy bands $(\pm E_2(\vec k))$ touch only at the
$K$-points where they have a quadratic dispersion. Finite $t_3\neq 0$
causes a splitting into in total four Dirac cones with linear
dispersion, an effect known as trigonal warping
\cite{McCann13,Mariani12}.

As a further important property, the eigenvectors corresponding to
$(\pm E_2(\vec k))$ are discontinuous as a function of wave vector at
the degeneracy points defined by Eq.~(\ref{cond2}); for more technical
details we refer to appendix \ref{continuity}. As a simplistic toy
model mimicking such an effect one can consider the Hamiltonian
$H=-k\sigma^z$ with a one-dimensional wave number $k$ and the Pauli
matrix $\sigma^z$ describing some internal degree of freedom: In the
many-body ground state of zero Fermi energy all occupied states with
$k>0$ have spin up while for all states with $k<0$ the spin points
downwards, resulting in a discontinuity of the occupied eigenvectors
at $k=0$.  As we shall see below, in the present case of graphene
bilayers this discontinuity is also reflected in the entanglement
spectrum.

An effective Hamiltonian providing an approximate description of the
central bands $(\pm E_2(\vec k))$ can be given following
Ref.~\onlinecite{McCann06}. In up to linear order in $1/t_{\perp}$ one
finds
\begin{small}
\begin{equation}
  H=-\left(
\begin{array}{cc}
  0 & 
  \frac{t^2}{t_{\perp}}\left(\gamma^{\ast}(\vec k)\right)^2+t_3\gamma(\vec k) \\
  \frac{t^2}{t_{\perp}}\left(\gamma(\vec k)\right)^2+t_3\gamma^{\ast}(\vec k) & 0
\end{array}
\right)
\label{hameff}
\end{equation}
\end{small}
with respect to the basis $\left(b_{2\vec k}^\dagger ,a_{1\vec
    k}^\dagger \right)|0\rangle$.  The eigenstates read
\begin{equation}
|\chi_{\pm}\rangle=\frac{1}{\sqrt{2}}\left(
\begin{array}{c}
  1 \\
  \mp e^{i\psi_{\vec k}}
\end{array}
\right)
\label{hameffeigenstate}
\end{equation}
with 
\begin{equation}
  e^{i\psi_{\vec k}}
  =\frac{\frac{t^2}{t_{\perp}}\left(\gamma(\vec k)\right)^2+t_3\gamma^{\ast}(\vec k)}
  {\left|\frac{t^2}{t_{\perp}}\left(\gamma(\vec k)\right)^2
      +t_3\gamma^{\ast}(\vec k)\right|}\,.
\end{equation}
Note that the Hamiltonian (\ref{hameff}) vanishes if and only if the
conditions (\ref{cond1}) or (\ref{cond2}) are fulfilled implying that
the positions of the central and satellite Dirac cones are the same as
for the full Hamiltonian (\ref{hambilayer}).  Moreover , $\psi_{\vec
  k}$ is a smooth and well-defined function of the wave vector except
for the locations of Dirac cones. Accordingly, the Berry curvature
\begin{equation}
  F(\vec k)=\frac{\partial A_y}{\partial k_x}
  -\frac{\partial A_x}{\partial k_y}
\label{berrycurv}
\end{equation}
arising from the Berry connection
\begin{equation}
  \vec A(\vec k)=i\langle\chi_{\pm}(\vec k)|
  \frac{\partial}{\partial\vec k}|\chi_{\pm}(\vec k)\rangle
  =-\frac{1}{2}\frac{\partial\psi_{\vec k}}{\partial\vec k}
\label{berrycon}
\end{equation}
vanishes everywhere outside the Dirac cones where contributions in
terms of $\delta$-functions arise. Integrating the Berry connection along
closed path in $\vec k$-space leads to geometrical quantities often
referred to as Berry phases, although no contact to adiabaticity is made here.
Moreover, if the Berry curvature has only nonzero contributions in terms
of $\delta$-functions (as it is the case here and in the following) these
geometrical phases are indeed topological, i.e. they are invariant under
continuous variations of the paths as long as the support of the
$\delta$-functions is not touched.

As discussed in
Refs.~\onlinecite{Manes07,Mikitik08,Mariani12}, integrating along a closed path
around the central Dirac cones at $K_{\pm}$ yields a Berry phase of $(\mp\pi)$,
while each of the
accompanying satellite cones gives a contribution of $(\pm\pi)$. Thus,
the total Berry phase arising at and around each $K$-point is, as in
the absence of trigonal warping, $(\pm 2\pi)$, and the integral over
the whole Brillouin zone of the Berry connection (i.e. the Chern
number) vanishes.
Naturally, our present analysis going beyond the Dirac cone
approximation confirms these results.
\section{Entanglement Spectra}
\label{entanglement}
For systems of free fermions as studied here, the entanglement
Hamiltonian can be formulated as a single-particle operator
\cite{Peschel03,Cheong04,Schliemann13},
\begin{equation}
  H_{\rm ent}=\sum_{\lambda}\xi_{\lambda}d^\dagger_{\lambda}d_{\lambda}\,.
\end{equation}
Here the $d^\dagger_{\lambda}$ generate eigenstates of the correlation
matrix
\begin{equation}
  C_{\alpha\beta}
  =\langle\Psi|c_{\alpha}^\dagger c_{\beta}|\Psi\rangle\,,
\end{equation}
where $|\Psi\rangle$ is the ground state of the composite system, and
single-particle operators $c_{\alpha}$, $c_{\beta}$ act on its
remaining part after tracing out a subsystem. The entanglement levels
$\xi_{\lambda}$ are related to the eigenvalues $\eta_{\lambda}$ of the
correlation matrix via
\begin{align}
\xi_{\lambda}=\ln\left(\frac{1-\eta_{\lambda}}{\eta_{\lambda}}\right)
={}& 2\artanh\left(1-2\eta_{\lambda}\right)\,.
\end{align}
In particular, the entanglement Hamiltonian and the correlation matrix share
the same system of eigenvectors.
\subsection{Tracing out One Layer}
\label{onelayer}
\begin{figure}[t]
  \includegraphics[width=\columnwidth]{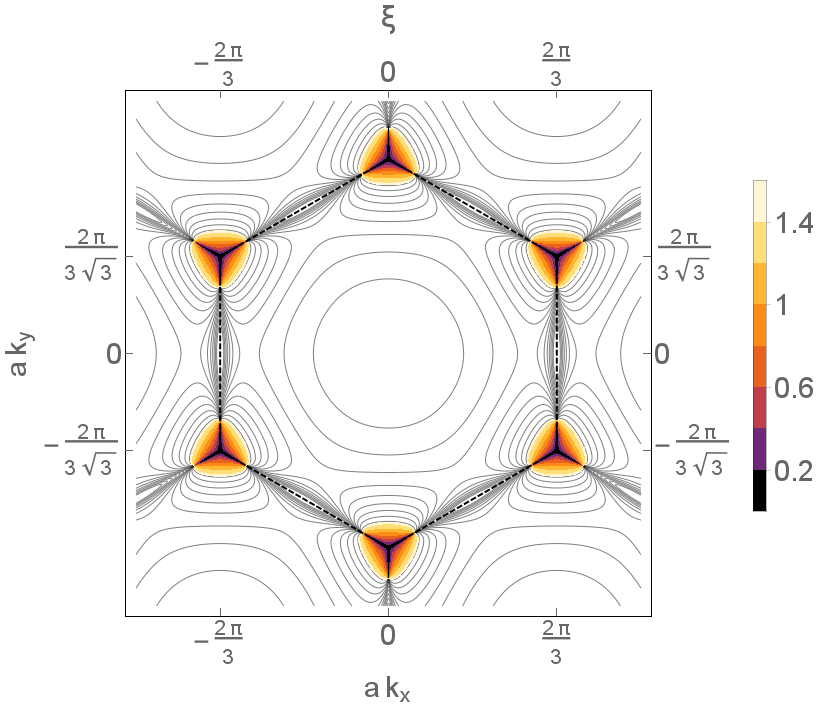}
  \caption{(Color Online) Contour plot of the entanglement spectrum
    $\xi_+(\vec k)$ plotted for $t_{\perp}=t$, $t_3=0.5t$. The contour
    of the colored region indicates $\xi=1.5$. The dashed line
    delineates the first Brillouin zone.}
\label{fig4}
\end{figure}

\begin{figure}[t]
  \includegraphics[width=\columnwidth]{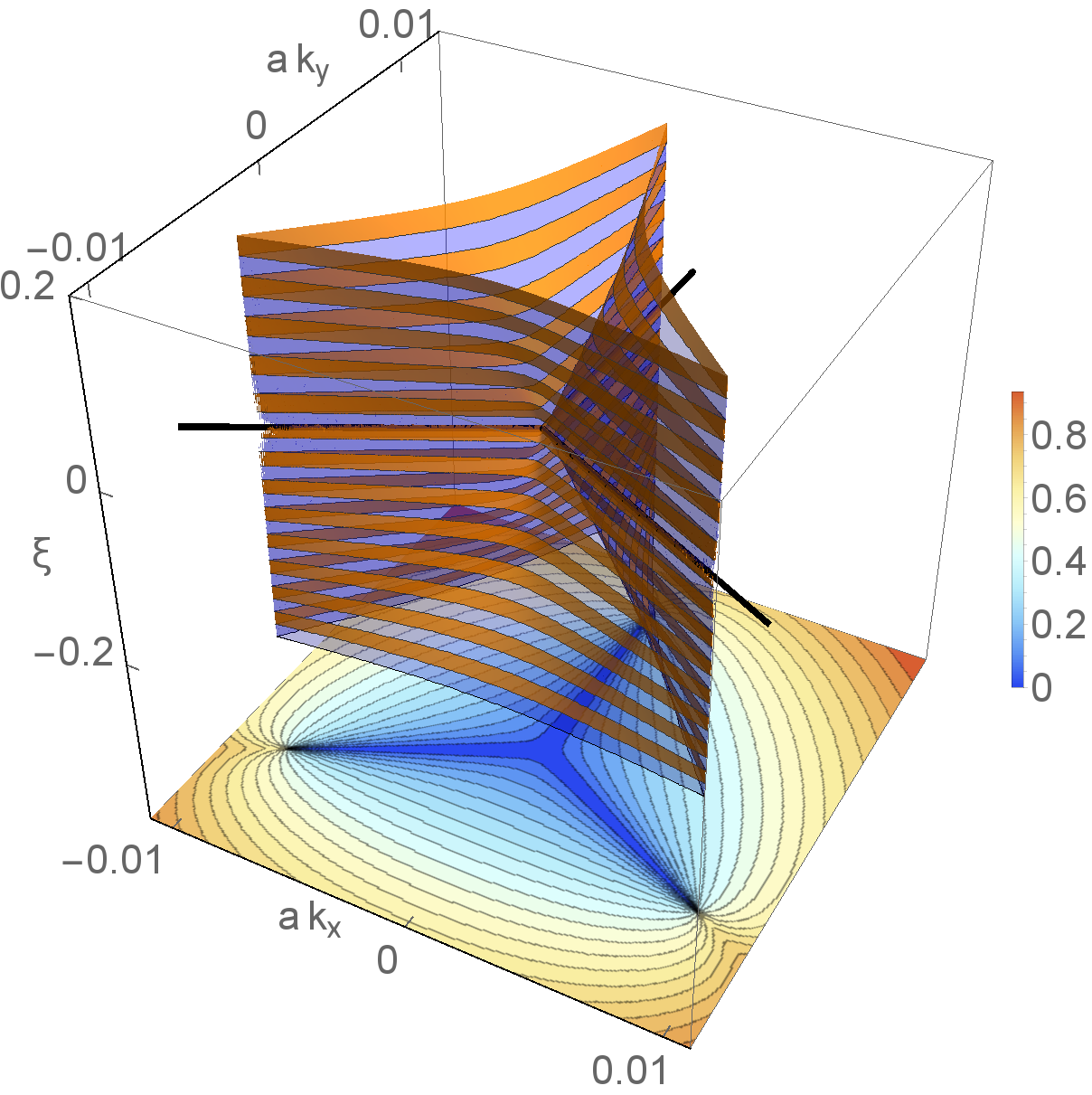}
  \caption{(Color Online) 
    The entanglement spectrum (\ref{entlevel}) plotted around a
    given $K$-point for the same parameters as in Fig.~\ref{fig3}.
    The density plot shows the upper entanglement level. Zero
    eigenvalues of the entanglement Hamiltonian occur along lines
    connecting the $K$-point with the locations of satellite Dirac
    cones of the energy spectrum (thick black lines).
    The components of the wave vector are measured relatively to the $K$-point.}
\label{fig5}
\end{figure}

We now consider the ground state of the undoped graphene bilayer such
that all states with negative energies $(-E_1(\vec k))$, 
$(-E_2(\vec k))$ are occupied while all others are empty. Tracing out layer 1
leads to the correlation matrix
\begin{equation}
  C(\vec k)=\left(
\begin{array}{cc}
  \frac{1}{2} & u(\vec k) \\
  u^{\ast}(\vec k) & \frac{1}{2}
\end{array}
\label{corrmat}
\right)
\end{equation}
where an explicit expression for $u(\vec k)$ is given in appendix
\ref{correlation}. The entanglement levels corresponding to the
eigenvalues $\eta_{\pm}(\vec k)=1/2\mp|u(\vec k)|$ are
\begin{align}
  \xi_{\pm}(\vec k) ={}& \pm 2\artanh\left(2|u(\vec k)|\right)\,.
\label{entlevel}
\end{align}
The modulus $|u|$ can be formulated as
\begin{equation}
  |u|=\frac{1/2}{\sqrt{1+(d/(t|\gamma(\vec k)|))^2}}
  \sqrt{\frac{1}{2}\left(1-\frac{\epsilon_1\epsilon_2+b^2}{E_1E_2}\right)}
\label{modu}
\end{equation}
with (cf. Eqs.~(\ref{defdapp}),(\ref{defbapp}))
\begin{eqnarray}
  d & = & \frac{\left(t_{\perp}^2-t_3^2|\gamma(\vec k)|^2\right)/2}
  {\sqrt{t_{\perp}^2+t_3^2|\gamma(\vec k)|^2
      -2t_{\perp}t_3|\gamma(\vec k)|\cos\left(3\phi_{\vec k}\right)}}\,,\\
  b & = & \frac{t_{\perp}t_3|\gamma(\vec k)||\sin\left(3\phi_{\vec k}\right)|}
  {\sqrt{t_{\perp}^2+t_3^2|\gamma(\vec k)|^2
      -2t_{\perp}t_{3}|\gamma(\vec k)|\cos\left(3\phi_{\vec k}\right)}}\,,
\end{eqnarray}
and (cf. Eq.~(\ref{defepsilon}))
\begin{small}
\begin{eqnarray}
  & & \epsilon_{1,2}=t|\gamma(\vec k)|\nonumber\\
  & & \pm\sqrt{\left(t_{\perp}^2+t_3^2|\gamma(\vec k)|^2
      -2t_{\perp}t_3|\gamma(\vec k)|\cos\left(3\phi_{\vec k}\right)\right)^2/4+d^2}
  \nonumber\\
\end{eqnarray}
\end{small}
implying
\begin{equation}
E_{1,2}=\sqrt{\epsilon_{1,2}^2+b^2}\,.
\end{equation}
The r.h.s of Eq.~(\ref{modu}) becomes zero if the radicand
vanishes. According to the discussion in appendices \ref{continuity}
and \ref{correlation} this is the case when 
$\cos\left(3\phi_{\vec k}\right)=-1$ leading to $b=0$ and $E_1=\epsilon_1\geq 0$,
$E_2=|\epsilon_2|$ such that
\begin{equation}
  |u|\propto\sqrt{\frac{1}{2}\left(1-\frac{\epsilon_2}{|\epsilon_2|}\right)}
\end{equation}
Now equation (\ref{epsiloncond}) shows that $|u(\vec k)|=0$ is
equivalent to
\begin{equation}
  \cos\left(3\phi_{\vec k}\right)=-1\quad
  \wedge\quad|\gamma(\vec k)|\in\left[0,t_{\perp}t_3/t^2\right]\,,
\label{cond3}
\end{equation}
where the endpoint of the above interval defines according to
condition (\ref{cond2}) the location of the satellite Dirac cones.  As
a result, the entanglement levels (\ref{entlevel}) vanish along
segments of the faces of the first Brillouin zone bounded by the
positions of the central Dirac cones and their satellites. At the
satellite Dirac cones the entanglement spectrum is discontinuous as a
function of wave vector. In Fig.~\ref{fig4} we plotted the
entanglement spectrum $\xi_+(\vec k)$ for the whole Brillouin
zone. For a better visualization large hopping parameters have been
chosen. The contour of the colored region connects all three satellite
Dirac cones. As discussed in appendix \ref{continuity}, this
discontinuity is inherited from a discontinuity in the eigenvectors of
the occupied single-particle states. The entanglement spectrum 
in the entire Brillouin zone is illustrated in Fig.~\ref{fig4}, whereas
Fig.~\ref{fig5} focuses on a given $K$-point.

Moreover, apart from the eigenvalues of the entanglement Hamiltonian,
let us also consider its eigenvector which coincide with the
eigenvectors of the correlation matrix (\ref{corrmat}). As discussed
in appendix \ref{correlation}, the complex function $u(\vec k)$
entering the correlation matrix becomes singular at the $K$-points and
the positions of the accompanying satellite Dirac cones of the energy
spectrum, leading again to $\delta$-function-type contributions to the
Berry curvature which vanishes otherwise.  Combining symbolic computer
algebra techniques and numerical calculations we find here a
Berry phase of $(\mp\pi/2)$ around the corners $K_{\pm}$ of the
Brillouin zone, and $(\pm\pi/2)$ for the corresponding satellite
positions.  For the central positions the above calculations can also
be done fully analytically by expanding the eigensystem data around
$K_{\pm}$.  For the satellite locations such an expansion is not
possible due to the discontinuity of the eigenvectors.

Thus, the total Berry phase contribution from each
$K$-point $K_{\pm}$ is $(\pm\pi)$ and agrees with the Berry phase
around the Dirac cones in monolayer graphene.  As a result, although
the entanglement spectrum of graphene bilayers generated by tracing
out one layer shows obvious differences to the energy spectrum of
monolayer graphene regarding qualitative geometrical properties, the
topological Berry phases obtained from the corresponding eigenvectors
still coincide at each $K$-point.
\subsection{Tracing out other Sublattices}

\begin{figure}[t]
  \includegraphics[width=\columnwidth]{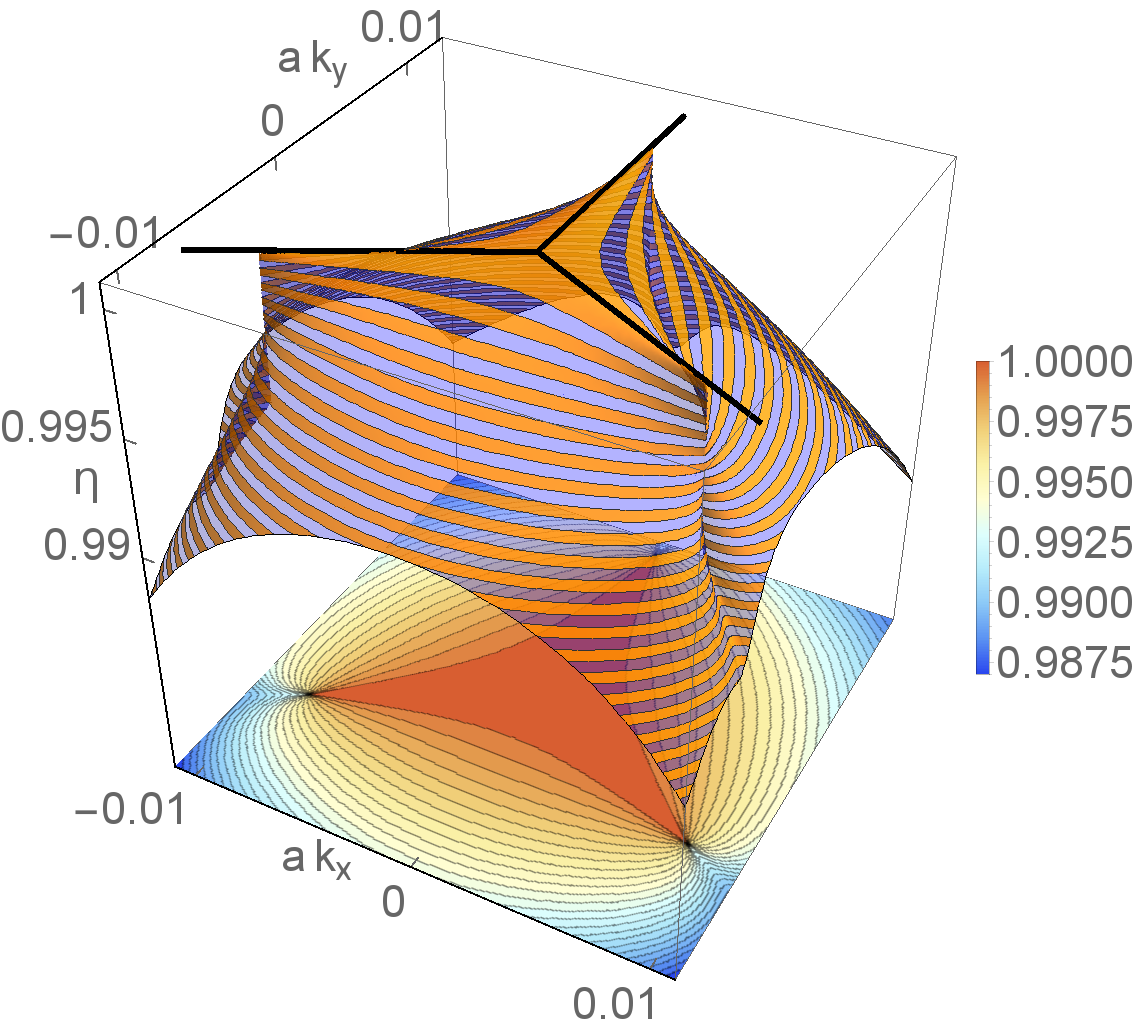}
  \caption{(Color Online) Eigenvalues $\eta_-(\vec k)=1/2+|v(\vec k)|$ of the
    correlation matrix plotted around a given $K$-point for
    $t_{\perp}=0.1t$, $t_3=0.15t$. The thick black lines correspond to
    the one in Fig.~\ref{fig5}, and the components of the wave vector are 
    again measured relatively to the $K$-point.}
\label{fig6}
\end{figure}

Now, we will consider the entanglement spectrum obtained by tracing out
sublattices A1 and B2 (or A2 and B1) lying in different layers.
In the former case one finds
\begin{equation}
  C(\vec k)=\left(
\begin{array}{cc}
  \frac{1}{2} & v(\vec k) \\
  v^{\ast}(\vec k) & \frac{1}{2}
\end{array}
\label{corrmat2}
\right)
\end{equation}
where an explicit expression for $v(\vec k)$ is given in appendix
\ref{correlation}. 
The above correlation matrix has eigenvalues 
$\eta_{\pm}(\vec k)=1/2\mp|v(\vec k)|$ leading to the entanglement levels 
\begin{align}
\xi_{\pm}(\vec k) ={}& \pm 2 \artanh \left(2|v(\vec k)|\right)\,.
\label{entlevel2}
\end{align}
In Fig.~\ref{fig6} we plotted the eigenvalues 
$\eta_-(\vec k)=1/2+|v(\vec k)|$ of the
correlation matrix around a given $K$-point.
The modulus $|v(\vec k)|$ reads more explicitly
\begin{align}
|v(\vec k)| ={}& \frac{1}{2}\sqrt{1
-\frac{t^2|\gamma(\vec k)|^2}{t^2|\gamma(\vec k)|^2+d^2}
\frac{1}{2}\left(1-\frac{\epsilon_1\epsilon_2+b^2}{E_1E_2}\right)}\\
 ={}& \frac{1}{2}\sqrt{1-4|u(\vec k)|^2}
\end{align}
and has a similar structure as $|u(\vec k)|$ given in Eq.~(\ref{modu}).
In particular, $|v(\vec k)|=1/2\Leftrightarrow|u(\vec k)|=0$ if the
conditions (\ref{cond3}) are fulfilled. In this case $\eta_+=0$ and
$\eta_-=1$ indicating that the remaining subsystem is unentangled with the
system traced out. 

Regarding Berry phases generated from the eigenstates of the correlation matrix
(\ref{corrmat2}) we note that the off-diagonal element $v(\vec k)$ nowhere
vanishes. As a consequence the Berry curvature defined analogously as
in Eqs.~(\ref{hameffeigenstate})-(\ref{berrycon}) is zero throughout the
Brillouin zone, which in turn holds for all Berry phases.
The nonvanishing of $v(\vec k)$ follows from the fact that
$|v(\vec k)|=0$ would require $|u(\vec k)|=1/2$ such that the
entanglement (\ref{entlevel}) would diverge which is, as seen in section
\ref{onelayer}, not the case.

Finally, the correlation matrix obtained by tracing over the sublattices
A1, A2 (or B1, B2) is proportional to the unit matrix,
\begin{align}
  C(\vec k) ={}& \left(
\begin{array}{cc}
  \frac{1}{2} & 0 \\
  0 & \frac{1}{2}
\end{array}
\label{corrmat3}
\right),
\end{align}
indicating that these sublattices are maximally entangled with the part
traced out.
\section{Conclusions and Outlook}
\label{concl}

We have studied entanglement properties of the ground state of Bernal
stacked graphene bilayers in the presence of trigonal warping. Our
analysis includes both the eigenvalues of the reduced density matrix
(giving rise to the entanglement spectrum) as well as its
eigenvectors.  When tracing out one layer, the entanglement spectrum
shows qualitative geometric differences to the energy spectrum of a
graphene monolayer while topological quantities such as Berry phase
type contributions to Chern numbers agree.  The latter finding is in
contrast to the reduced density matrix resulting from tracing out
other sublattices of the bilayer system.  Here, all corresponding
Berry phase integrals yield trivially zero.  Thus, our study provides
an example for common topological properties of the eigensystem of the
energy Hamiltonian of a subsystem (here a graphene monolayer) and the
entanglement Hamiltonian, while the geometrical shape of both spectra
grossly differs.  Our investigations are based on closed analytical
expressions for the full eigensystem of bilayer graphene in the entire
Brillouin zone with a trigonally warped spectrum.

Future work might address bilayer systems of other geometrical
structures such as the Kagome lattice, the influence of a static
perpendicular magnetic field \cite{Nemec07,Schliemann13}, and the
effect of time-periodic in-plane electric fields \cite{Wang15}.

{\em Note added.} After this paper was made available as an arXiv preprint
and submitted to the journal, we became aware of Ref.~\cite{Fukui14}
where also Chern numbers calculated from the eigenstates of entanglement
Hamiltonians are studied. Most recent work building upon this concept
is reported on in Ref.~\cite{Araki16}.

\section*{Acknowledgments}

This work was supported by Deutsche Forschungsgemeinschaft via GRK1570.

\appendix
\begin{widetext}
\section{Diagonalization of the Bilayer Hamiltonian}
\label{diag}
Putting $t_4=0$ and fixing a wave vector $\vec k$ the Hamiltonian
(\ref{hambilayer}) reads with respect to the basis $\left(a_{2\vec
    k}^\dagger,b_{1\vec k}^\dagger,b_{2\vec k}^\dagger,a_{1\vec k}^\dagger\right)|0\rangle$
\begin{equation}
H=\left(
\begin{array}{cccc} 
  0 & t_{\perp} & -t \gamma(\vec k) & 0 \\
  t_{\perp} & 0 & 0 & -t \gamma^{\ast}(\vec k) \\
  -t \gamma^{\ast}(\vec k) & 0 & 0 & -t_3 \gamma(\vec k) \\
  0 & -t \gamma(\vec k) & -t_3 \gamma^{*}(\vec k) & 0 \\
\end{array}
\right)\,.
\label{hambilayer2}
\end{equation} 
Using $\gamma(\vec k)=|\gamma(\vec k)|e^{i\phi_{\vec k}}$ we apply the
transformation
\begin{equation}
U_{1}=\frac{1}{\sqrt{2}}\left(
\begin{array}{cccc}
  1 & 1 & 0 & 0 \\
  0 & 0 & e^{i\phi_{\vec{k}}} & e^{-i\phi_{\vec{k}}} \\
  0 & 0 & e^{i\phi_{\vec{k}}} & -e^{-i\phi_{\vec{k}}} \\
  1 & -1 & 0 & 0 \\
\end{array}
\right)
\end{equation}
such that in
\begin{equation}
H_1=U_1HU_1^\dagger=\left(
\begin{array}{cccc}
  t_{\perp} & -t|\gamma(\vec k)| & 0 & 0 \\
  -t|\gamma(\vec k)| 
  & -t_3|\gamma(\vec k)|\cos\left(3\phi_{\vec k}\right) 
  & it_3|\gamma(\vec k)|\sin\left(3\phi_{\vec k}\right) & 0 \\
  0 & -it_3|\gamma(\vec k)|\sin\left(3\phi_{\vec k}\right) 
  & t_3|\gamma(\vec k)|\cos\left(3\phi_{\vec k}\right) 
  & -t|\gamma(\vec k)| \\
  0 & 0 & -t|\gamma(\vec k)|& -t_{\perp}
\end{array}
\right)
\end{equation}
all information on the phase $\phi_{\vec k}$ is contained in the matrix elements
being proportional to the skew parameter $t_3$. Proceeding now with
the transformation
\begin{equation}
U_2=\frac{1}{\sqrt{2}}\left(
\begin{array}{cccc}
  1 & -1 & 0 & 0 \\
  1 & 1 & 0 & 0 \\
  0 & 0 & 1 & -1 \\
  0 & 0 & 1 & 1 
\end{array}
\right)
\end{equation}
we find
\begin{equation}
  H_2=U_2H_1U_2^\dagger=\frac{1}{2}
  \left(
\begin{array}{cccc}
  e_1 & c & -is & -i s \\
  c & e_2 & is & is \\
  is & -is  & -e_2 & c \\
  is & -is & c & -e_1 
\end{array}
\right)
\end{equation}
with
\begin{eqnarray}
  e_1 & = & \phantom{-}2t|\gamma(\vec k)| + t_{\perp}
  -t_3|\gamma(\vec k)|\cos\left(3\phi_{\vec k}\right) \,,\\
  e_2 & = & -2t|\gamma(\vec k)| + t_{\perp}
  -t_3|\gamma(\vec k)|\cos\left(3\phi_{\vec k}\right)\,,\\
  c & = & t_{\perp}+t_3|\gamma(\vec k)|\cos\left(3\phi_{\vec k}\right)\,,\\
  s & = & t_3|\gamma(\vec k)|\sin\left(3\phi_{\vec k}\right)\,.
\end{eqnarray}
Here it is useful to split the above matrix as $ H_2=H_2^{'}+H_2^{''} $
where
\begin{equation}
H_2^{'}=\frac{1}{2}\left(
\begin{array}{cccc}
  e_1 & 0 & -is & 0 \\
  0 & e_2 & 0 & is \\
  is & 0  & -e_2 & 0 \\
  0 & -is & 0 & -e_1 
\end{array}
\right)
\quad,\quad
H_2^{''}=\frac{1}{2}\left(
\begin{array}{cccc}
  0 & c & 0 & -i s \\
  c & 0 & is & 0 \\
  0 & -is  & 0 & c \\
  is & 0 & c & 0 
\end{array}
\right)\,.
\end{equation}
$H_2^{'}$ is diagonalized by
\begin{equation}
U_3=\left(
\begin{array}{cccc} 
  \alpha_+ & 0 & -i\sigma\alpha_- & 0 \\
  0 & -i\sigma\alpha_+ & 0 & \alpha_- \\
  -i\sigma\alpha_- & 0 & \alpha_+ & 0 \\
  0 & \alpha_- & 0 & -i\sigma\alpha_+
\end{array}
\right) 
\end{equation}
with $\sigma = \sign\left(\sin(3\phi(\vec k))\right)$ and
\begin{align}
  \alpha_{\pm} ={}& \sqrt{\frac{1}{2}\left(1\pm\frac{t_{\perp}-
        t_3|\gamma(\vec k)|\cos\left(3\phi_{\vec k}\right)}
      {\sqrt{t_{\perp}^2+t_3^2|\gamma(\vec k)|^2
          -2t_{\perp}t_{3}|\gamma(\vec k)|\cos\left(3\phi_{\vec
              k}\right)}}\right)}
\end{align}
such that
\begin{equation}
  H_3=U_3H_2U_3^\dagger=\left(
\begin{array}{cccc}
  \zeta_1 & id\sigma & 0 & b \\
  -id\sigma & \zeta_2 & b & 0 \\
  0 & b & -\zeta_2 & id\sigma \\
  b & 0 & -id\sigma & -\zeta_1 
\end{array}
\right)
\end{equation}
where
\begin{eqnarray}
  d & = & \frac{\left(t_{\perp}^2-t_3^2|\gamma(\vec k)|^2\right)/2}
  {\sqrt{t_{\perp}^2+t_3^2|\gamma(\vec k)|^2
      -2t_{\perp}t_3|\gamma(\vec k)|\cos\left(3\phi_{\vec k}\right)}}\,,
\label{defdapp}\\
b & = & \frac{t_{\perp}t_3|\gamma(\vec k)||\sin\left(3\phi_{\vec k}\right)|}
{\sqrt{t_{\perp}^2+t_3^2|\gamma(\vec k)|^2
-2t_{\perp}t_{3}|\gamma(\vec k)|\cos\left(3\phi_{\vec k}\right)}}\,,
\label{defbapp}
\end{eqnarray}
and $\pm\zeta_1$ and $\pm\zeta_2$ are eigenvalues  of $H_2^{'}$ given by
\begin{equation}
\zeta_{1/2}=\frac{1}{2}\left(\pm 2t|\gamma(\vec k)|
+\sqrt{t_{\perp}^2+t_3^2|\gamma(\vec k)|^2-2t_{\perp}t_3|\gamma(\vec k)|
\cos\left(3\phi_{\vec k}\right)}\right)\,.
\end{equation}
Splitting now $H_3$ in the form
\begin{equation}
H_3=\left(
\begin{array}{cccc}
  \zeta_1 & id & 0 & 0 \\
  -id & \zeta_2 & 0 & 0 \\
  0 & 0 & -\zeta_2 & id \\
  0 & 0 & -id & -\zeta_1 
\end{array}
\right)
+\left(
\begin{array}{cccc}
  0 & 0 & 0 & b \\
  0 & 0 & b & 0 \\
  0 & b & 0 & 0 \\
  b & 0 & 0 & 0 
\end{array}
\right)
\end{equation}
the first part is diagonalized by
\begin{eqnarray}
U_4=\left(
\begin{array}{cccc}
  -i\sigma\tau\beta_+  & \beta_- & 0 & 0 \\
  \beta_- & -i\sigma\tau\beta_+ & 0 & 0 \\
  0 & 0 & -i\sigma\tau\beta_+ & \beta_- \\
  0 & 0 & \beta_- & -i\sigma\tau\beta_+  
\end{array}
\right)
\end{eqnarray}
with $\tau = \sign(d)$ and 
\begin{align}
  \beta_{\pm} = \sqrt{
    \frac{1}{2}
    \left( 1 \pm \frac{\zeta_1-\zeta_2}{
      \sqrt{(\zeta_1-\zeta_2)^2+4d^2}}
    \right)
  }
\end{align}
while the second part is left unchanged by $U_4$ resulting in
\begin{equation}
H_4=U_4H_3U_4^\dagger=\left(
\begin{array}{cccc}
  \epsilon_1 & 0 & 0 & b \\
  0 & \epsilon_2 & b & 0 \\
  0 & b & -\epsilon_2 & 0 \\
  b & 0 & 0 &  -\epsilon_1
\end{array}
\right)
\end{equation}
with the diagonal elements are given in terms of
\begin{equation}
\epsilon_{1/2}=\frac{1}{2}\left(\zeta_1+\zeta_2
\pm\sqrt{\left(\zeta_1-\zeta_2\right)^2+4d^2}\right)\,.
\label{defepsilon}
\end{equation} 
Finally, $H_4$ is brought into diagonal form via
\begin{equation}
U_5=\left(
\begin{array}{cccc}
  \gamma_+^{(1)} & 0 & 0 & \gamma_-^{(1)} \\
  0 & \gamma_+^{(2)} & \gamma_-^{(2)} & 0 \\
  0 & \gamma_-^{(2)} & -\gamma_+^{(2)} & 0 \\
  \gamma_-^{(1)} & 0 & 0 & -\gamma_+^{(1)}
\end{array}
\right)
\end{equation}
with
\begin{equation}
\gamma_{\pm}^{(1)}=\sqrt{\frac{1}{2}\left(1\pm\frac{\epsilon_1}{E_2}\right)}
\quad,\quad 
\gamma_{\pm}^{(2)}=\sqrt{\frac{1}{2}\left(1\pm\frac{\epsilon_2}{E_2}\right)}
\label{defgamma}
\end{equation}
and
\begin{eqnarray}
  & & E_{1/2}=\sqrt{\epsilon_{1,2}^2+b^2}\\
  &  & =\sqrt{\frac{1}{2}\left(t_{\perp}^2+t_3^2|\gamma(\vec k)|^2
      +2t^2|\gamma(\vec k)|^2
      \pm\sqrt{4t^2|\gamma(\vec k)|^2\left(t_{\perp}^2+t_3^2|\gamma(\vec k)|^2
          -2t_{\perp}t_3|\gamma(\vec k)|\cos\left(3\phi_{\vec k}\right)\right)
        +\left(t_{\perp}^2-t_3^2|\gamma(\vec k)|^2\right)^2}\right)}.\nonumber\\
\end{eqnarray}
Thus,
\begin{align}
U_5H_4U_5^\dagger ={}& \diag\left(E_1,E_2,-E_2,-E_1\right)\,,
\end{align}
and the matrix elements of the corresponding total transformation
$U=U_5U_4U_3U_2U_1$ can be expressed as
\begin{eqnarray}
  U_{11} & = & \frac{1}{2}\left(\alpha_--i\sigma\alpha_+\right)
  \left(\tau\beta_++\beta_-\right)\left(\gamma_+^{(1)}-i\sigma\gamma_-^{(1)}\right)\\
  U_{12} & = & \frac{1}{2}\left(\alpha_+-i\sigma\alpha_-\right)
  \left(\tau\beta_++\beta_-\right)\left(\gamma_-^{(1)}-i\sigma\gamma_+^{(1)}\right)\\
  U_{13} & = & -\frac{e^{i\phi_{\vec k}}}{2}\left(\alpha_--i\sigma\alpha_+\right)
  \left(\tau\beta_+-\beta_-\right)\left(\gamma_+^{(1)}+i\sigma\gamma_-^{(1)}\right)\\
  U_{14} & = & \frac{e^{-i\phi_{\vec{k}}}}{2}\left(\alpha_-+i\sigma \alpha_+\right)
  \left(\tau\beta_+-\beta_-\right)\left(\gamma_+^{(1)}-i\sigma\gamma_-^{(1)}\right)
\end{eqnarray}
and 
\begin{eqnarray}
  U_{21} & = & -\frac{1}{2}\left(\alpha_++i\sigma\alpha_-\right)
  \left(\tau\beta_+-\beta_-\right)\left(\gamma_+^{(2)}-i\sigma\gamma_-^{(2)}\right)\\
  U_{22} & = & -\frac{1}{2}\left(\alpha_+-i\sigma \alpha_-\right)
  \left(\tau\beta_+-\beta_-\right)\left(\gamma_+^{(2)}+i\sigma\gamma_-^{(2)}\right)\\
  U_{23} & = & -\frac{e^{i\phi_{\vec k}}}{2}\left(\alpha_++i\sigma \alpha_-\right)
  \left(\tau\beta_++\beta_-\right)\left(\gamma_+^{(2)}+i\sigma\gamma_-^{(2)}\right)\\
  U_{24} & = & -\frac{e^{-i\phi_{\vec k}}}{2}\left(\alpha_+-i\sigma \alpha_-\right)
  \left(\tau\beta_++\beta_-\right)\left(\gamma_+^{(2)}-i\sigma\gamma_-^{(2)}\right)
\end{eqnarray}
which are the complex conjugates of the components of the eigenvectors
of the conduction-band states with positive energies $E_1(\vec k)$,
$E_2(\vec k)$, while
\begin{eqnarray}
  U_{31} & = & \frac{1}{2}\left(\alpha_--i\sigma \alpha_+\right)
  \left(\tau\beta_+-\beta_-\right)\left(\gamma_+^{(2)}-i\sigma\gamma_-^{(2)}\right)\\
  U_{32} & = & \frac{1}{2}\left(\alpha_-+i\sigma \alpha_+\right)
  \left(\tau\beta_+-\beta_-\right)\left(\gamma_+^{(2)}+i\sigma\gamma_-^{(2)}\right)\\
  U_{33} & = & -\frac{e^{i\phi_{\vec k}}}{2}\left(\alpha_++i\sigma \alpha_-\right)
  \left(\tau\beta_++\beta_-\right)\left(\gamma_-^{(2)}-i\sigma\gamma_+^{(2)}\right)\\
  U_{34} & = & -\frac{e^{-i\phi_{\vec k}}}{2}\left(\alpha_+-i\sigma \alpha_-\right)
  \left(\tau\beta_++\beta_-\right)\left(\gamma_-^{(2)}+i\sigma\gamma_+^{(2)}\right)
\end{eqnarray}
and
\begin{eqnarray}
  U_{41} & = & \phantom{-}\frac{1}{2}\left(\alpha_++i\sigma \alpha_-\right)
  \left(\tau\beta_++\beta_-\right)\left(\gamma_+^{(1)}-i\sigma\gamma_-^{(1)}\right)\\
  U_{42} & = & -\frac{1}{2}\left(\alpha_-+i\sigma \alpha_+\right)
  \left(\tau\beta_++\beta_-\right)\left(\gamma_-^{(1)}-i\sigma\gamma_+^{(1)}\right)\\
  U_{43} & = & \frac{e^{i\phi_{\vec k}}}{2}\left(\alpha_++i\sigma \alpha_-\right)
  \left(\tau\beta_+-\beta_-\right)\left(\gamma_+^{(1)}+i\sigma\gamma_-^{(1)}\right)\\
  U_{44} & = & \frac{e^{-i\phi_{\vec{k}}}}{2}\left(\alpha_-+i\sigma \alpha_+\right)
  \left(\tau\beta_+-\beta_-\right)\left(\gamma_-^{(1)}+i\sigma\gamma_+^{(1)}\right)
\end{eqnarray}
correspond to the valence-band states with negative energies
$(-E_2(\vec k))$, $(-E_1(\vec k))$. Note that all factors involving
$\alpha_{\pm}$, $\gamma_{\pm}^{(1)}$, $\gamma_{\pm}^{(2)}$ in the
above expressions have modulus one, i.e. they are phase factors.
\section{Continuity Properties}
\label{continuity}
The eigenvectors corresponding to the energy branches $(\pm E_2(\vec
k))$ are discontinuous at wave vectors determined by the condition
(\ref{cond2}).  This comes about as follows: The matrix elements
$U_{2,n}(\vec k)$, $U_{3,n}(\vec k)$, $n\in\{1,2,3,4\}$ contain the
quantities $\gamma_{\pm}^{(2)}$ defined in Eqs.~(\ref{defgamma})
whereas the $U_{1,n}(\vec k)$, $U_{4,n}(\vec k)$ corresponding to
$(\pm E_1(\vec k))$ involve $\gamma_{\pm}^{(1)}$ Fixing now
$\cos\left(\phi_{\vec k}\right)=-1$ we have $b=0$ such that
$E_1=\epsilon_1\geq 0$ and $E_2=|\epsilon_2|$ such that
$\gamma_{\pm}^{(1)}$ remain continuous while $\gamma_{\pm}^{(2)}$
become
\begin{equation}
  \gamma_{\pm}^{(2)}=\sqrt{\frac{1}{2}
    \left(1\pm\frac{\epsilon_2}{|\epsilon_2|}\right)}\,.
\end{equation}
Inspection of Eq.~(\ref{defepsilon}) now shows that for
$\cos\left(\phi_{\vec k}\right)=-1$
\begin{equation}
\epsilon_2(\vec k)\left\{
\begin{array}{cc}
  >0 & \,|\gamma(\vec k)|<t_{\perp}t_3/t^2 \\
  <0 & \,|\gamma(\vec k)|>t_{\perp}t_3/t^2 
\end{array}\right.
\label{epsiloncond}
\end{equation}
such that $\epsilon_2(\vec k)$ changes sign for $|\gamma(\vec
k)|=t_{\perp}t_3/t^2$, i.e. $\gamma_{\pm}^{(2)}$ is discontinuous at
wave vectors given by the condition (\ref{cond2}). This discontinuity
is inherited by the correlation matrix and, in turn, by the
entanglement spectrum.

The technical reason for this discontinuity in the eigenvectors is of
course the fact that the dispersions $(\pm E_2(\vec k))$ become
degenerate at wave vectors fulfilling (\ref{cond2}).  In fact the
eigenvectors can also be considered as continuous functions of the
wave vector by appropriately relabeling the dispersion branches. In
the ground state of the undoped bilayer system, however, only the
lower branch $(-E_2(\vec k))$ is occupied, which makes the
discontinuity unavoidable.

To circumvent this discontinuity one can open an energy gap between
the upper and lower central band such that the corresponding
eigenstates are necessarily continuous for all wave vectors. Among the
various mechanisms producing such a gap only few allow for a still
halfway convenient analytical treatment of the Hamiltonian. These
include introducing identical mass terms in both layers,
i.e. $H\mapsto H+H^{\prime}$ with
\begin{align}
  H^{\prime} ={}& \diag\left(m,-m,-m,m\right)\,,
\end{align}
or applying a bias voltage $\Lambda$ between the layers,
\begin{align}
  H^{\prime} ={}& \diag\left(-\Lambda/2,\Lambda/2,-\Lambda/2,\Lambda/2\right)\,.
\end{align}
In the former case the four dispersion branches $(\pm E_1(\vec k))$,
$(\pm E_2(\vec k))$ are given by
\begin{eqnarray}
  E_{1/2}(\vec k) & = & 
  \Biggl[m^2+\frac{1}{2}\left(t_{\perp}^2+t_3^2|\gamma(\vec k)|^2
    +2t^2|\gamma(\vec k)|^2\right)\nonumber\\
  & & \pm\frac{1}{2}
  \sqrt{4t^2|\gamma(\vec k)|^2\left(t_{\perp}^2+t_3^2|\gamma(\vec k)|^2
      -2t_{\perp}t_3|\gamma(\vec k)|\cos\left(3\phi_{\vec k}\right)\right)
    +\left(t_{\perp}^2-t_3^2|\gamma(\vec k)|^2\right)^2}\Biggr]^{1/2}
\label{ergspecm}
\end{eqnarray}
while for a bias voltage one finds \cite{McCann06}
\begin{eqnarray}
  E_{1/2}(\vec k) & = & 
  \Biggl[\frac{\Lambda^2}{4}+\frac{1}{2}\left(t_{\perp}^2+t_3^2|\gamma(\vec k)|^2
    +2t^2|\gamma(\vec k)|^2\right)\nonumber\\
  & & \pm\frac{1}{2}
  \sqrt{4t^2|\gamma(\vec k)|^2\left(t_{\perp}^2+t_3^2|\gamma(\vec k)|^2
      -2t_{\perp}t_3|\gamma(\vec k)|\cos\left(3\phi_{\vec k}\right)+\Lambda^2\right)
    +\left(t_{\perp}^2-t_3^2|\gamma(\vec k)|^2\right)^2}\Biggr]^{1/2}\,.
\label{ergspecU}
\end{eqnarray}
In both cases the central energy bands $(\pm E_2(\vec k))$ are
separated by a gap, and the spectrum can still be given in terms of
comparably simple closed expressions since the characteristic
polynomial of the $4\times 4$ Hamiltonian matrix is a second-order
polynomial in the energy squared leading to a spectrum being symmetric
around zero. Also the corresponding eigenvectors can be obtained in
closed analytical forms by procedures analogous to (but in detail
somewhat more complicated than) the one given in appendix \ref{diag}
\cite{Predin15}.

Note that applying a bias voltage as well as introducing a mass term
in each layer discriminates the layers against each other. The latter
circumstance is due to the fact that $t_{\perp}$ couples sublattices
in different layers for which the mass term has different sign. As a
result, when tracing out, say, one layer of an undoped
(i.e. half-filled) bilayer system, the remaining layer will not be
half-filled, what obscures somewhat the comparison with an undoped
graphene monolayer.
\section{Correlation Matrices}
\label{correlation}
Upon tracing out layer $1$ from the ground state of the undoped
bilayer system the correlation matrix reads in the basis
$\left(a_{2\vec k}^\dagger,b_{2\vec k}^\dagger\right)|0\rangle$
\begin{equation}
C(\vec k)=\left(
\begin{array}{cc}
  U_{31}U_{31}^{\ast}+U_{41}U_{41}^{\ast} & U_{31}U_{33}^{\ast}+U_{41}U_{43}^{\ast} \\
  U_{33}U_{31}^{\ast}+U_{43}U_{41}^{\ast} & U_{33}U_{33}^{\ast}+U_{43}U_{43}^{\ast} 
\end{array}
\right)
=\left(
\begin{array}{cc}
\frac{1}{2} & u(\vec k) \\
u^{\ast}(\vec k) & \frac{1}{2}
\end{array}
\right)
\end{equation}
with
\begin{equation}
  u(\vec k)=\frac{e^{-i\phi_{\vec k}}}{4}
  \left(\beta_+^2-\beta_-^2\right)
  \left(\left(\gamma_+^{(1)}-i\sigma\gamma_-^{(1)}\right)^2
    -\left(\gamma_+^{(2)}-i\sigma\gamma_-^{(2)}\right)^2\right)\,.
\label{u}
\end{equation}
This quantity becomes singular at the corners of the Brillouin zone
where $\gamma(\vec k)$ is zero such that its phase is ill-defined, and
at the positions of the satellite Dirac cones of the energy spectrum
where, as discussed in appendix \ref{continuity}, $\gamma_{\pm}^{(2)}$
is discontinuous.

Tracing out the sublattices A1 and B2 one finds in the basis
$\left(a_{2\vec k}^\dagger,b_{1\vec k}^\dagger\right)|0\rangle$
\begin{equation}
C(\vec k)=\left(
\begin{array}{cc}
  U_{31}U_{31}^{\ast}+U_{41}U_{41}^{\ast} & U_{31}U_{32}^{\ast}+U_{41}U_{42}^{\ast} \\
  U_{32}U_{31}^{\ast}+U_{42}U_{41}^{\ast} & U_{32}U_{32}^{\ast}+U_{42}U_{42}^{\ast} 
\end{array}
\right)
=\left(
\begin{array}{cc}
  \frac{1}{2} & v(\vec k) \\
  v^{\ast}(\vec k) & \frac{1}{2}
\end{array}
\right)
\end{equation}
with
\begin{equation}
  v(\vec k)=\frac{\left(\alpha_--i\sigma\alpha_+\right)^2}{4}
  \left(\left(\tau\beta_+-\beta_-\right)^2
    \left(\gamma_+^{(2)}-i\sigma\gamma_-^{(2)}\right)^2
    +\left(\tau\beta_++\beta_-\right)^2
    \left(\gamma_+^{(1)}-i\sigma\gamma_-^{(1)}\right)^2\right)\,.
\label{v}
\end{equation}
Note that the expressions (\ref{u}),(\ref{v}) obey the interesting sum rule
\begin{equation}
  |u(\vec k)|^2+|v(\vec k)|^2=\frac{1}{4}
\end{equation}
which s fulfilled whenever the coefficients involved satisfy
\begin{equation}
\alpha_+^2+\alpha_-^2=\beta_+^2+\beta_-^2
=\left(\gamma_+^{(1/2)}\right)^2+\left(\gamma_-^{(1/2)}\right)^2=1\,,
\end{equation}
which is the case here by construction.

Finally, the correlation matrix obtained by tracing out the sublattices
A1, A2 is proportional to the unit matrix,
\begin{equation}
  C(\vec k)=\left(
\begin{array}{cc}
  U_{32}U_{32}^{\ast}+U_{42}U_{42}^{\ast} & U_{32}U_{33}^{\ast}+U_{42}U_{43}^{\ast} \\
  U_{33}U_{32}^{\ast}+U_{43}U_{42}^{\ast} & U_{33}U_{33}^{\ast}+U_{43}U_{43}^{\ast} 
\end{array}
\right)
=\left(
\begin{array}{cc}
  \frac{1}{2} & 0 \\
  0 & \frac{1}{2}
\end{array}
\right)
\end{equation}
implying that the remaining subsystem is maximally entangled with
the subsystem traced out.

\end{widetext}
{}
\end{document}